\documentstyle[11pt,emulateapj,epsf,rotate]{article} 

\lefthead{Popov et al.}
\righthead{LOG N-LOG S DISTRIBUTIONS OF ISOLATED NEUTRON STARS}

\begin{document}
\title {Log N -- Log S distributions of accreting and cooling isolated
neutron stars}

\author{S.B. Popov\altaffilmark{1}, M. Colpi\altaffilmark{2},
M.E. Prokhorov\altaffilmark{1}, A. Treves\altaffilmark{3}
and R. Turolla\altaffilmark{4}}
\altaffiltext{1}{Sternberg Astronomical Institute, Universitetskii Pr. 13,
119899, Moscow, Russia; e--mail: polar@sai.msu.ru}
\altaffiltext{2}{Dipartimento di Fisica, Universit\`a di Milano Bicocca,
P.zza
della Scienza 3, 20126 Milano, Italy; e--mail: colpi@uni.mi.astro.it}
\altaffiltext{3}{Dipartimento di Scienze, Universit\`a  dell'Insubria,
Via Lucini 3,
22100, Como, Italy; e--mail: treves@mi.infn.it}
\altaffiltext{4}{Dipartimento di Fisica, Universit\`a di Padova, Via
Marzolo 8,
35131 Padova,
Italy; e--mail: turolla@pd.infn.it}

\begin{abstract}

We model populations of isolated neutron stars in the Galaxy
following their orbital and magneto-rotational evolution.
It is shown that accretors become more abundant than coolers
at fluxes below $\sim 10^{-13}$ erg cm$^{-2}$ s$^{-1},$
and one can predict that about one accreting neutron star per square 
degree should be observed at the {\it Chandra} and {\it Newton} flux limits of
$\sim 10^{-16}$ erg cm$^{-2}$ s$^{-1}.$
The soft ROSAT sources associated with isolated neutron stars
can be relatively young cooling objects only if the neutron star birth rate
in the Solar vicinity during the last $\sim 10^6$ yr is higher than
that inferred from radiopulsar observations.

\end{abstract}

\noindent

\section{Introduction}

Despite intensive observational campaigns, no irrefutable identification of an isolated accreting neutron star (NS) has been presented so far. Six soft 
sources have been found in ROSAT fields which are most 
probably associated to  isolated radioquiet NSs. 
Present X-ray and optical data however do not allow an unambiguous
identification of the physical mechanism responsible for their emission. 
These sources
can be powered either by accretion of the interstellar gas onto old ($\approx
10^{10}$ yr) NSs or by the release of internal energy in relatively young ($\approx
10^6$ yr) cooling NSs (see Treves et al. \cite{t2000} and Motch 
\cite{motch2000} for recent reviews). The ROSAT
candidates, although relatively bright (up to $\approx 1 \ {\rm ct\,s}^{-1}$), are
intrinsically dim and their inferred luminosity ($L\approx 10^{31} \ {\rm erg\, s}^{-1}$)
is near to that expected from either a close-by cooling NS or 
from an accreting NS among the most luminous.
Their X-ray spectrum
is soft and thermal, again as predicted for both accretors and coolers (Zane, Turolla \&
Treves \cite{zzt2000}; Treves et al. \cite{t2000}).
Up to now only two optical counterparts have been identified (RXJ 1856,
Walter \& Matthews \cite{wm97}, for which a distance estimate of $\sim 60$ 
pc has been very recently obtained, Walter \cite{w2000}, and RXJ 0720, 
Kulkarni \& Van Kerkwick \cite{kk98}). In both cases an
optical excess over the low-frequency tail of the black body X-ray spectrum 
has been reported.
While detailed multiwavelength observations with next-generation instruments may indeed be
the key for assessing the true nature of these sources, other, indirect, approaches may be
used to discriminate in favor of one of the two scenarios proposed so far.

In this {\it Letter}  we compute and compare the $\log N$ -- $\log S$ distribution of both
accreting and cooling NSs, to establish the relative
contribution of the two populations to the observed number counts.
Previous studies derived the $\log N$ -- $\log S$
distribution of accretors (Treves \& Colpi \cite{tc91}; Madau and Blaes \cite{mb94};
Manning et al. \cite{mann96}) assuming  a NSs velocity distribution rich in slow
stars ($v\lesssim 100 \ {\rm km\, s}^{-1}$). More recent measurements of pulsar
velocities (e.g. Lyne \& Lorimer \cite{ll94}; Hansen \& Phinney \cite{hp97})
and upper limits on the observed number of accretors in ROSAT surveys (Danner
\cite{dan98a}, \cite{dan98b}) point, however, to a larger NS mean velocity
(see Treves et al. \cite{t2000} for a critical discussion).
Recently Ne\"uhauser \& Tr\"umper (\cite{nt99}, NT99 hereafter)
compared the number count distribution of the
ROSAT isolated NS  candidates  with those of accretors and coolers.
Here we address these issues in  greater detail, also
in the light of the latest contributions to the modeling of the evolution
of Galactic NSs (Popov et al. \cite{p2000}, P2000 hereafter).
In $\S 2$ we use a population synthesis model, developed in P2000, to trace the
properties of the NS spin, magnetic field and
accretion luminosity in the Galactic potential using a detailed map of
the interstellar medium (ISM).
Cooling NSs are explored separately in $\S 3 $ within a simpler model of local
sources. Results are discussed in $\S 4$.

\section{Accreting Isolated Neutron Stars}

The census model adopted here follows closely that developed in P2000.
All NSs are born in the Galactic plane,
their birth rate is taken to be proportional to the
square of the ISM density and is constant in time.
Initially a NS has a circular
velocity fixed by its position in the Galactic plane and determined by the
Galactic potential.  To this, an
additional kick velocity with random orientation and modulus selected from
a Maxwellian distribution is imparted.

Time resolution is adapted to follow in detail the different evolutionary phases
the star experiences. NSs are born in the ejector (E) phase and then, as
braking due to electromagnetic torques increases the period, 
come to the propeller (P)
phase. Further spin-down, produced by accretion torques, 
may drive the star in the
accretion (A) phase when matter can penetrate the magnetosphere 
reaching the stellar
surface (see e.g. Lipunov \cite{l92}).
Accretion is then assumed to proceed at the local Bondi rate
and the large scale ISM distribution is mapped as in P2000. 
The small scale  structure of the ISM ($\approx 100$AU-1 pc, see e.g. 
Meyer \& Lauroesh \cite{ml99}; Treves et al. \cite{t2000}), 
has not been taken into 
account as it would produce an unacceptable increase of the 
computing time, leading to a worse statistics. A detailed mapping
of the ISM on all scales is fundamental in 
assessing the observability of single, luminous sources (see Zane et 
al. \cite{za96}), but the effects produced by its multiphase structure 
on the evolution of the entire NS population are likely to average out.

Here we consider only the simplest picture, where the star magnetic field
is constant and the dipole moment $\mu$ has a log-gaussian
distribution, similar to that observed in radiopusars
($\langle\log\mu\rangle =30.06$, $\sigma =0.32$ in CGS units).

The magneto-rotational evolution is followed using essentially the same approach as
in P2000. In particular, the typical values of the transition periods are $\sim 25$ s
and $\sim 1500$ s  for E$\to$ P and  P$\to$ A transitions, respectively.
As soon as the star enters the accretor stage, the period is set to
the value typical of an accretor embedded in a turbulent medium
(Lipunov \& Popov \cite{lp95}; Konenkov \& Popov \cite{kp97}).

The statistics of accretors was improved
by a factor $>10$ relative to P2000 by
evolving only stars from the low-velocity tail ($v<100$ km s$^{-1}$) of a
Maxwellian with $\langle v\rangle = 300$ km s$^{-1}$.
This assumption appears justified since the time spent in the
ejector phase increases with velocity,
$t_E\sim 10^{10}\, \mu_{30}^{-1}n^{-1/2}
v_{100} \, {\rm yr}\, . $
The present choice for $\langle v\rangle$ follows from radiopulsar observations and
current limits on the detection of accreting NSs (P2000).
Results can be easily scaled to any $\langle v\rangle > 200$ km s$^{-1}$.

Nearly all highly magnetized isolated NSs in the velocity range
considered here become accretors, and in calculating the
$\log N$ -- $\log S$ distribution we  obtain good statistics for fluxes $<10^{-11}$
erg cm$^{-2}$ s$^{-1}$. At larger fluxes the number of sources is too small to
give reliable counts, but the extrapolation $\log N\sim -(3/2)\log S$ can be 
safely used for $\log S\gtrsim -11$.

Our main results are presented in Figures \ref{fig1} and \ref{fig3}, and
refer to a total of
$10^9$ NSs in the Galaxy. The $\log N$ -- $\log S$
distribution
of accretors is computed for sources within a distance of 5 kpc,
together with the
distributions of velocity, accretion rate and effective
temperature (calculated
adopting as effective surface the polar cap area).
Isolated NSs with fields $<0.5\times  10^{11}$ G
never appear as accretors in any run.
The brightest accretors have luminosities  $\sim 10^{32}$ erg
s$^{-1}$, but for the majority of them $L\approx 10^{29}-10^{30}$ erg s$^{-1}$.
Our sources are relatively hot, with a mean effective temperature 300-400 eV
and for them absorption can be neglected.
We predict, on average, about 1 source per square
degree, in the energy range  0.5-2 keV, for limiting fluxes about
$10^{-16}$--$10^{-15}$ erg cm$^{-2}$ s$^{-1}$. This implies $\sim 3\times 10^4$
sources in the whole sky. Note that they are significantly concentrated
toward the Galactic plane with a  center-anticenter asymmetry.

\medskip
\epsfxsize=7.6truecm
\rotate[r]{\epsfbox{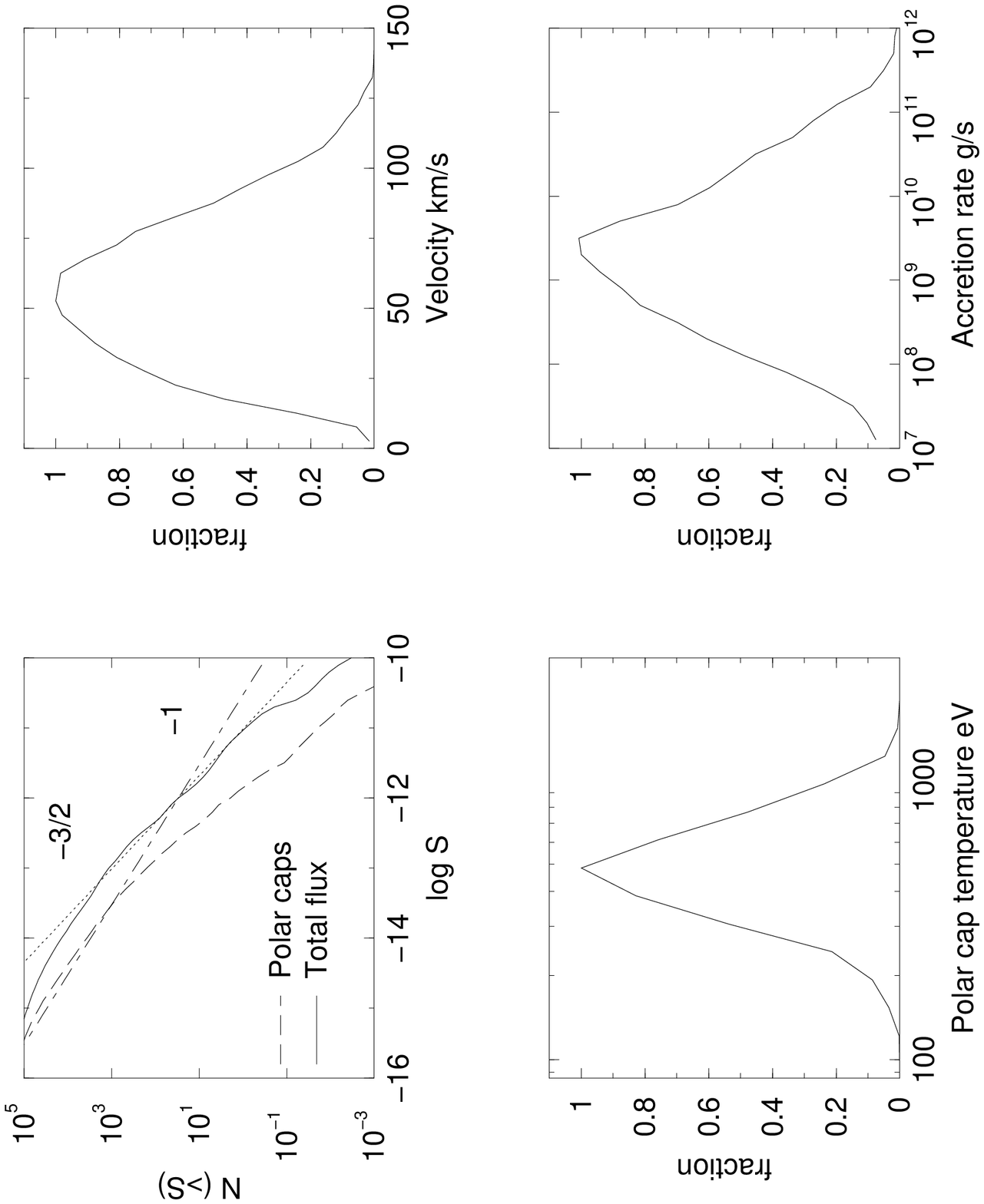}}
\figcaption[fig1.ps]{\label{fig1}
\small{Upper left panel: the $\log N$ -- $\log S$ distribution for
accretors within 5 kpc from the Sun. The two curves refer to emission
from the entire star surface and
to polar cap emission in the range 0.5-2 keV; two straight lines with slopes -1 and -3/2
are also shown for comparison. From top right to bottom right: the velocity,
effective temperature and accretion rate distributions of accretors; all
distributions are normalized to their maximum value.
}}
\medskip
\newpage
\section{Cooling Neutron Stars}

Soon after their birth in a type II supernova event, 
cooling NSs emit X-ray luminosities as high as 
 $\sim 10^{34}$ erg s$^{-1}$ for a few thousands years.
Before disappearing from the X-ray sky, they 
shine as faint
soft sources ($L \sim 10^{32}$ erg s$^{-1}$) for about a million year
(e.g. Yakovlev et al. \cite{yak99}).

In the determination of the log$N$-log$S$, 
the cooler sample should in principle be drawn from the same parent population 
as accretors. However, as their total number is quite low ($\approx 10^{-4}$
of the total number of NS in the Galaxy) owing to 
the short duration of the cooling phase, 
the results would be affected by the very poor statistics.
Notice however that coolers can travel only a distance $\approx 400$ pc during
their lifetime and have softer spectra than accretors,
so they appear
mostly as a local population of sources (due to interstellar absorption).

For this reason, we prefer to approximate their spatial distribution around the sun as
homogeneous with a scale height of $\sim 450$ pc.  We take cooling
NSs as ``standard candles'' with $L=10^{32}$ erg s$^{-1}$, giving off a black-body
spectrum with effective temperature $\sim 50\,$ eV.  
The duration of the cooling phase is taken to be $10^6$
yr, according to the ``slow cooling'' scenario (see NT99).  We used two typical values
for the total NS spatial density: $0.33\times 10^{-3}\, {\rm pc}^{-3}$ and $3.3\times
10^{-3}\, {\rm pc}^{-3}$.  The former estimate follows from radiopulsar statistics and
was used by NT99, while the latter corresponds to a total population of $10^9$ NSs, as
implied by constraints on nucleosynthetic yields and historic supernova rates (Arnett
et al. \cite{ar89}; van den Bergh \& Tammann \cite{vdbt91}).  Accordingly, coolers have
a mean density of $0.33 - 3.3\times 10^{-7} {\rm pc}^{-3}.$
 
The ISM structure is treated in a very simple way. The ISM density inside
the Local Bubble of radius $r_l=140 \ {\rm pc}$ (Sfeir et al. \cite{sfeir99})
is  $n=0.07$ cm$^{-3}$. At larger distances the interstellar gas
is assumed to be uniformly distributed in a disk of half-thickness 450 pc
with $n=1$ cm$^{-3}$.

Results are shown in Figure \ref{fig2}, where we compare the observed
distribution of ROSAT candidates
with that of coolers, as predicted by the present model (including absorption) and by
NT99, the latter
obtained for a total NS number   $N_{tot} = 2\times 10^8$.
As it is apparent, coolers dominate over accretors at large fluxes (for the same $N_{tot}$),
essentially because of their much higher luminosity ($10^{32}$ vs.
$10^{29}-10^{30}$ erg s$^{-1}$).

This simple model reproduces the most important feature, the ``flattening''
of the $\log N$ -- $\log S$ distribution outside the Local Bubble.
Such strong flattening can help to explain the observed data,
if one assumes that most of the brightest isolated NSs are identified
(Schwope et al. \cite{setal99}). 

The value of the flux at which the distribution flattens out (the ``knee'') depends on
the size of the Local Bubble $r_l.$ The ISM densities and the duration of the cooling
time affect instead the number of sources.  Since the cooling time can be an important
parameter, the comparison of theoretical and observed distributions could, in
principle, help in discriminating between ``fast'' and ``slow'' cooling models. 

\medskip
\epsfxsize=8truecm
\rotate[r]{\epsfbox{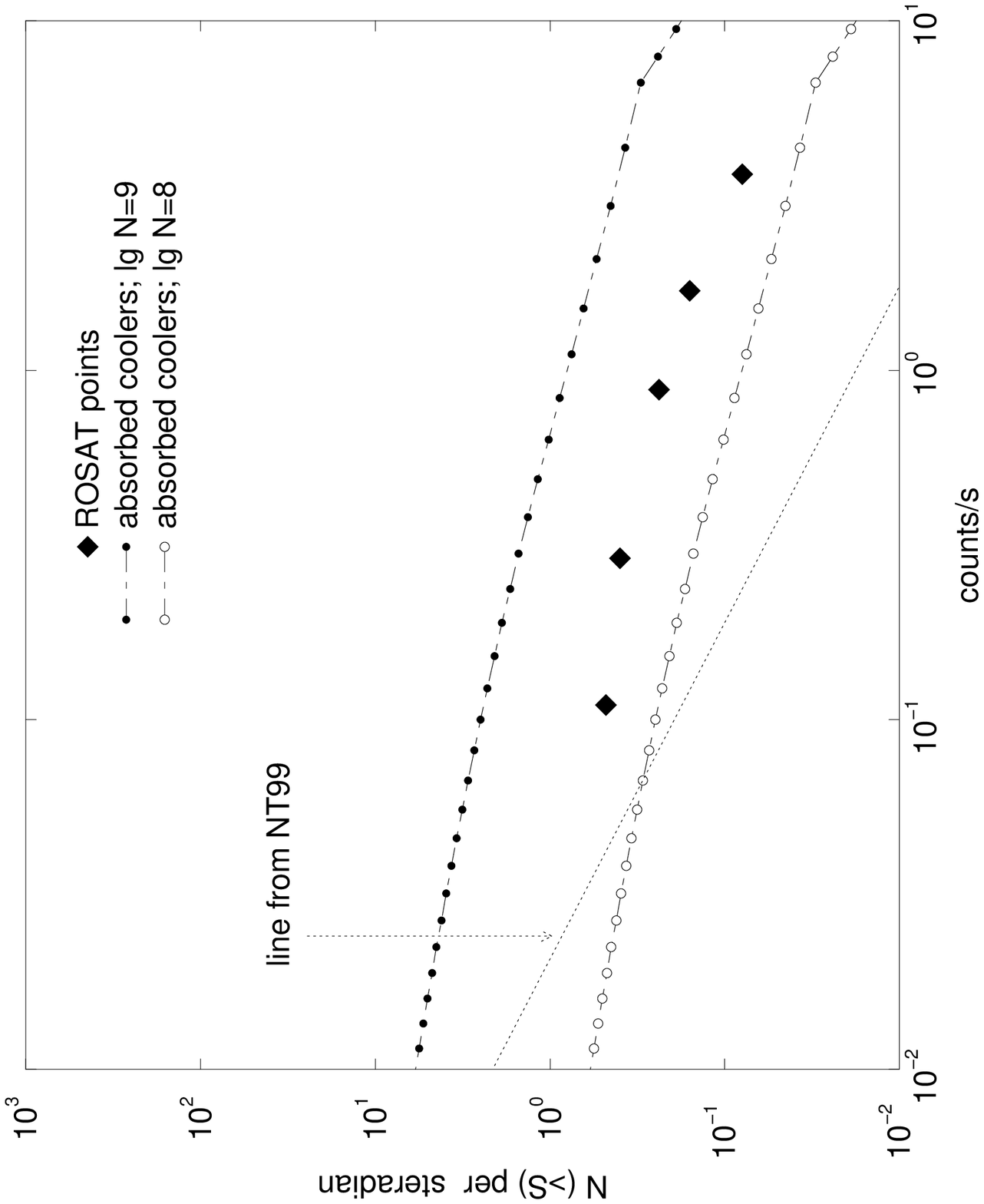}}
\figcaption[fig2.ps]{\label{fig2}
\small{The $\log N$--$\log S$ for coolers.
The upper curve (filled circles) correspond to a NS density of $3.3\times 10^{-3}$ pc$^{-3}$ 
and the lower one (open circles) to $3.3\times 10^{-4}$ pc$^{-3}$; here $r_l=140$ pc.}}
\medskip

\section{Discussion}

The statistical analysis presented in the previous sections shows that
both scenarios, accretors and coolers, have difficulties in explaining,
under usual assumptions,  
the observed properties of isolated NS ROSAT candidates.
The main problem is to produce enough sources that are, at the same time:
i) relatively bright (high count rates, $\gtrsim 0.1$ cts s$^{-1}$); 
ii) close (low absorption 
$N_H \lesssim 10^{20} {\rm cm}^{-2}$); iii) soft ($T_{eff}\sim 50 - 100$ eV); iv) 
slowly rotating (for two candidates, RX J0720 and RX J0420, the detected periods are
8.4 and 22.7 s). 

Polar cap accretion for a non-decaying magnetic field, $B\approx 10^{12}$ G,
cannot 
produce enough bright sources and the emitted  spectrum is likely to be harder, 
see Figure \ref{fig1}. Moreover, accretors are expected to have much
larger periods, $P\gtrsim 10^3$ 
s. On the other hand, relatively young cooling NSs can be spun down 
to periods $\sim 20$ s in $<10^6$ yrs (when they are still hot) only for 
$B\gtrsim 10^{14}$ G, since $P\sim 15 \,  B_{14}(t/10^6\, {\rm yr})^{1/2}\, {\rm s}$
for dipole losses.
If all NSs experience an initial radiopulsar phase,
their total number in the Galaxy would be about
$(1-3)\times 10^8$ (Lyne et al. \cite{ll98}). In this case there will be 
not   bright enough isolated NSs in the Solar vicinity in both scenarios.

If $N_{tot}\sim  10^9$, coolers can explain the large number of bright
close-by sources
and their typical temperatures (see Figure \ref{fig2}). 
However, the problem of reproducing the 
observed periods still remains, unless the progenitors of the ROSAT pulsating 
candidates are NSs born with long periods, which never passed through the 
pulsar phase; alternatively they could be
magnetars with ultra-high fields, as suggested by Heyl \& Kulkarni (\cite{hk98}).
If proved correct, the ``fast'' cooling
scenario, in which the duration of the active X-ray phase is shorter, 
would make again 
problematic to account for the observed bright sources. 

In the accretion scenario, increasing the total number of NSs is also a
need. As already shown by Livio et al. (\cite{lxf98})
and Colpi et al.
(\cite{elu98}), magnetic field decay can explain the 
low temperatures and periods. Decay can both increase and decrease
number of accreting isolated NSs (Popov, \& Prokhorov \cite{pp2000}). 
But a complete statistical analysis is still to come.

\medskip
\vbox{
\epsfxsize=8truecm
\rotate[r]{\epsfbox{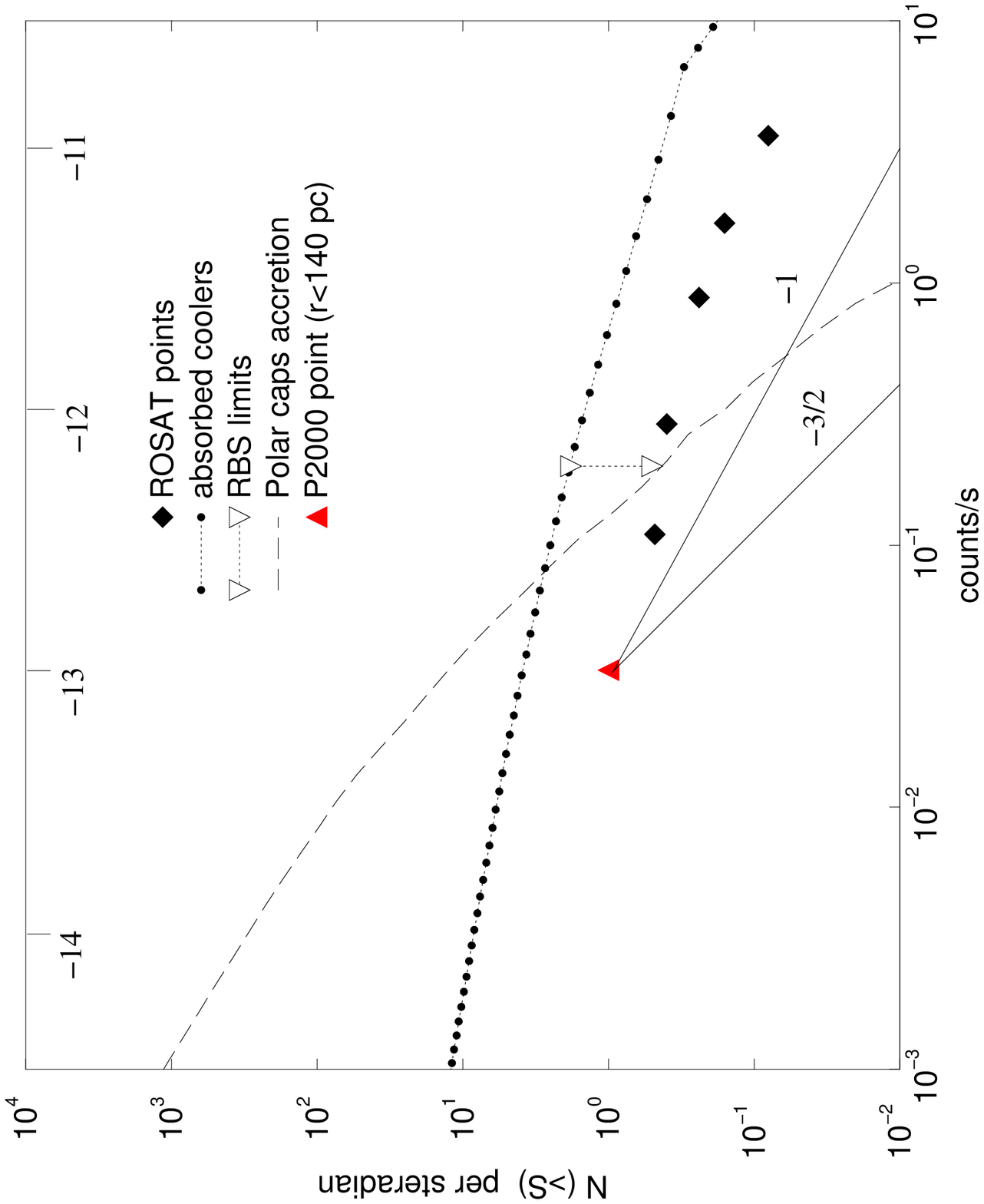}}
\figcaption[fig3.ps]{\label{fig3}
\small {Comparison of the log $N$ -- log $S$ distributions for accretors and coolers together
with observational points, the naive log$N$ -- log$S$ from
P2000 and the ROSAT Bright Survey (RBS) limit (Schwope et al. \cite{setal99}).
The scale on the top horizontal axes gives the flux in erg cm$^{-2}$s$^{-1}$.}}
}
\medskip

In any case, using  ``standard'' assumptions
on the velocity, spin period and magnetic field parameters,
the accretion scenario can not explain the observed properties of the six
ROSAT candidates.

Note that  the number of accretors (of all luminosities) in the Galaxy is
found to be about two orders of magnitude larger
than the number of young coolers:
few percents versus 0.01\% 
of the total number of isolated NSs.
But on average accretors are 3 orders of magnitude fainter than a typical
cooler. A key result of our  statistical analysis is that
accretors should  eventually become more abundant than coolers
at fluxes below
$10^{-13}$ erg cm$^{-2}$ s$^{-1}$. This is illustrated in Figure 
\ref{fig3}
in which we summarize the main outcome of our simulations.

\section*{Acknowledgments}
We wish to thank  J. Lattimer, E. van den Heuvel, V.M. Lipunov, and
S. Campana for useful discussions. SBP and MEP
also thank the University of Insubria for
financial support and the Universities of  Milano-Bicocca, Padova and
the Brera Observatory (Merate) for their kind hospitality.
The work of SBP and MEP was supported through grant RFBR 00-02-17164
and NTP Astronomy grants 1.4.4.1. and 1.4.2.3. Support from the European
Commission under contract ERBFMRXCT98-0195 and from MURST under contract
COFIN98021541 is acknowledged.

\end{document}